\documentstyle[12pt,epsfig]{article}

\newcommand{\agt}{\,\rlap{\lower 3.5 pt \hbox{$\mathchar \sim$}} \raise 1pt
 \hbox {$>$}\,}
\newcommand{\alt}{\,\rlap{\lower 3.5 pt \hbox{$\mathchar \sim$}} \raise 1pt
 \hbox {$<$}\,}

\catcode`@=11
\newcount\@tempcntc
\def\@citex[#1]#2{\if@filesw\immediate\write\@auxout{\string\citation{#2}}\fi
  \@tempcnta\z@\@tempcntb\m@ne\def\@citea{}\@cite{\@for\@citeb:=#2\do
    {\@ifundefined
       {b@\@citeb}{\@citeo\@tempcntb\m@ne\@citea\def\@citea{,}{\bf ?}\@warning
       {Citation `\@citeb' on page \thepage \space undefined}}%
    {\setbox\z@\hbox{\global\@tempcntc0\csname b@\@citeb\endcsname\relax}%
     \ifnum\@tempcntc=\z@ \@citeo\@tempcntb\m@ne
       \@citea\def\@citea{,}\hbox{\csname b@\@citeb\endcsname}%
     \else
      \advance\@tempcntb\@ne
      \ifnum\@tempcntb=\@tempcntc
      \else\advance\@tempcntb\m@ne\@citeo
      \@tempcnta\@tempcntc\@tempcntb\@tempcntc\fi\fi}}\@citeo}{#1}}
\def\@citeo{\ifnum\@tempcnta>\@tempcntb\else\@citea\def\@citea{,}%
  \ifnum\@tempcnta=\@tempcntb\the\@tempcnta\else
   {\advance\@tempcnta\@ne\ifnum\@tempcnta=\@tempcntb \else \def\@citea{--}\fi
    \advance\@tempcnta\m@ne\the\@tempcnta\@citea\the\@tempcntb}\fi\fi}
\catcode`@=12

\begin{document}

\title{\vskip-3cm{\baselineskip14pt
\centerline{\normalsize DESY 99-112\hfill ISSN 0418-9833}
\centerline{\normalsize MPI/PhT/99-33\hfill}
\centerline{\normalsize hep-ph/9908385\hfill}
\centerline{\normalsize August 1999\hfill}
}
\vskip1.5cm
$H^+H^-$ Pair Production at the Large Hadron Collider
}
\author{
{\sc A.A. Barrientos Bendez\'u}\\
{\normalsize II. Institut f\"ur Theoretische Physik, Universit\"at Hamburg,}\\
{\normalsize Luruper Chaussee 149, 22761 Hamburg, Germany}\\
\\
{\sc B.A. Kniehl}\thanks{Permanent address: II. Institut f\"ur
Theoretische Physik, Universit\"at Hamburg, Luruper Chaussee 149,
22761 Hamburg, Germany.}\\
{\normalsize Max-Planck-Institut f\"ur Physik (Werner-Heisenberg-Institut),}\\
{\normalsize F\"ohringer Ring 6, 80805 Munich, Germany}}

\date{}

\maketitle

\thispagestyle{empty}

\begin{abstract}
We study the pair production of charged Higgs bosons at the CERN Large Hadron
Collider in the context of the minimal supersymmetric extension of the
standard model.
We compare the contributions due to $q\bar q$ annihilation at the tree level
and $gg$ fusion, which proceeds at one loop.
At small or large values of $\tan\beta$, $H^+H^-$ production proceeds 
dominantly via $b\bar b$ annihilation, due to Feynman diagrams involving
neutral CP-even Higgs bosons and top quarks, which come in addition to the
usually considered Drell-Yan diagrams.
In the case of $gg$ fusion, the squark loop contributions may considerably 
enhance the well-known quark loop contributions.

\medskip

\end{abstract}

\newpage

\section{Introduction}

One of the prime objectives of the CERN Large Hadron Collider (LHC) is the
search for spin-zero particles which remain in the physical spectrum after the
elementary-particle masses have been generated through the Higgs mechanism of
electroweak symmetry breaking \cite{kun}.
Should the world be supersymmetric, then the Higgs sector is more complicated
than in the standard model (SM), which predicts just one scalar Higgs boson.
The Higgs sector of the minimal supersymmetric extension of the SM (MSSM) 
consists of a two-Higgs-doublet model (2HDM) and accommodates five physical
Higgs bosons:
the neutral CP-even $h^0$ and $H^0$ bosons, the neutral CP-odd $A^0$ boson,
and the charged $H^\pm$-boson pair.
At the tree level, the MSSM Higgs sector has two free parameters, which are
usually taken to be the mass $m_{A^0}$ of the $A^0$ boson and the ratio
$\tan\beta=v_2/v_1$ of the vacuum expectation values of the two Higgs
doublets.

The discovery of the $H^\pm$ bosons would prove wrong the SM and, at the same 
time, give strong support to the MSSM.
The logistics of the $H^\pm$-boson search at the LHC may be summarized as 
follows.
For $H^\pm$-boson masses $m_H<m_t-m_b$, the dominant production mechanisms are
$gg,q\bar q\to t\bar t$ followed by $t\to bH^+$ \cite{kun}.\footnote{Here and
in the following, the charge-conjugate processes will not be explicitly
mentioned.}
The dominant decay mode of $H^\pm$ bosons in this mass range is
$H^+\to \bar\tau\nu_\tau$ unless $\tan\beta<\sqrt{m_c/m_\tau}\approx1$
\cite{kun}.
In contrast to the SM top-quark events, this signature violates lepton
universality, a criterion which is routinely applied in ongoing $H^\pm$-boson
searches at the Fermilab $p\bar p$ collider Tevatron \cite{abe}.
For larger values of $m_H$, the most copious sources of $H^\pm$ bosons are 
provided by $g\bar b\to\bar tH^+$ \cite{gun}, $gg\to\bar tbH^+$ \cite{dia},
and $qb\to q^\prime bH^+$ \cite{mor}.
The preferred decay channel is then $H^+\to t\bar b$, independently of
$\tan\beta$ \cite{kun}.
An interesting alternative is to produce $H^\pm$ bosons in association with
$W^\mp$ bosons, so that the leptonic decays of the latter may serve as a
trigger for the $H^\pm$-boson search.
The dominant subprocesses of $W^\pm H^\mp$ associated production are
$b\bar b\to W^\pm H^\mp$ at the tree level and $gg\to W^\pm H^\mp$ at one 
loop, which were investigated for $m_b=0$ and small values of $\tan\beta$
($0.3\le\tan\beta\le2.3$) in Ref.~\cite{dic} and recently, without these
restrictions, in Ref.~\cite{wh}.
A careful signal-versus-background analysis, based on the analytic results of
Ref.~\cite{wh}, was recently reported in Ref.~\cite{smo}.

\begin{figure}[ht]
\begin{center}
\centerline{\epsfig{figure=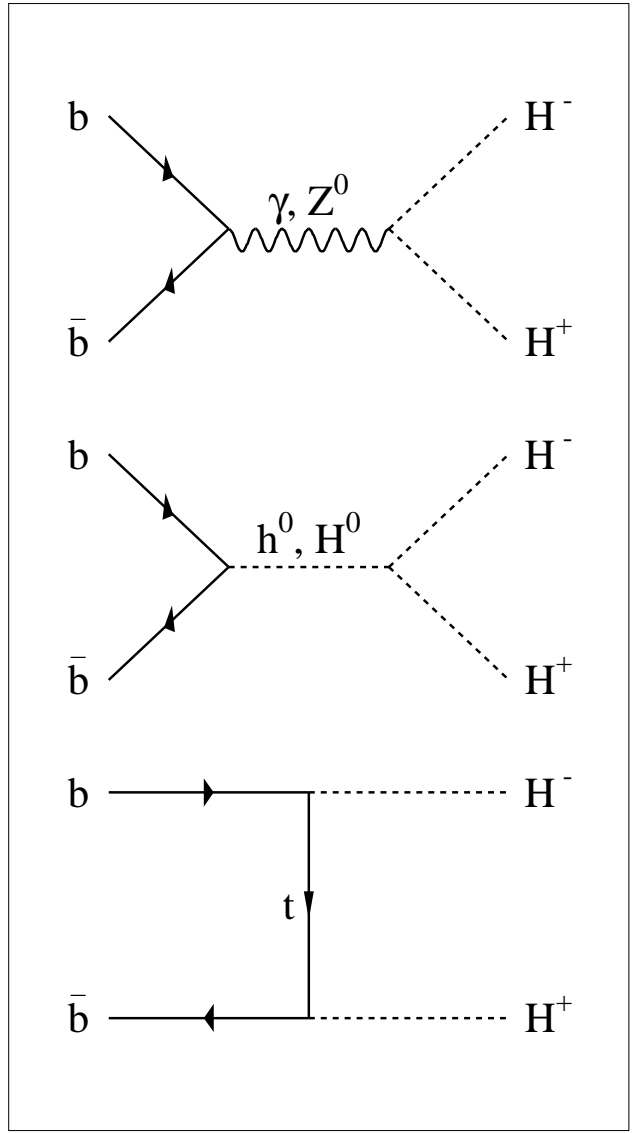,height=18cm}}
\caption{Tree-level Feynman diagrams for $b\bar b\to H^+H^-$ in the MSSM.}
\label{fig:one}
\end{center}
\end{figure}

\begin{figure}[ht]
\begin{center}
\centerline{\epsfig{figure=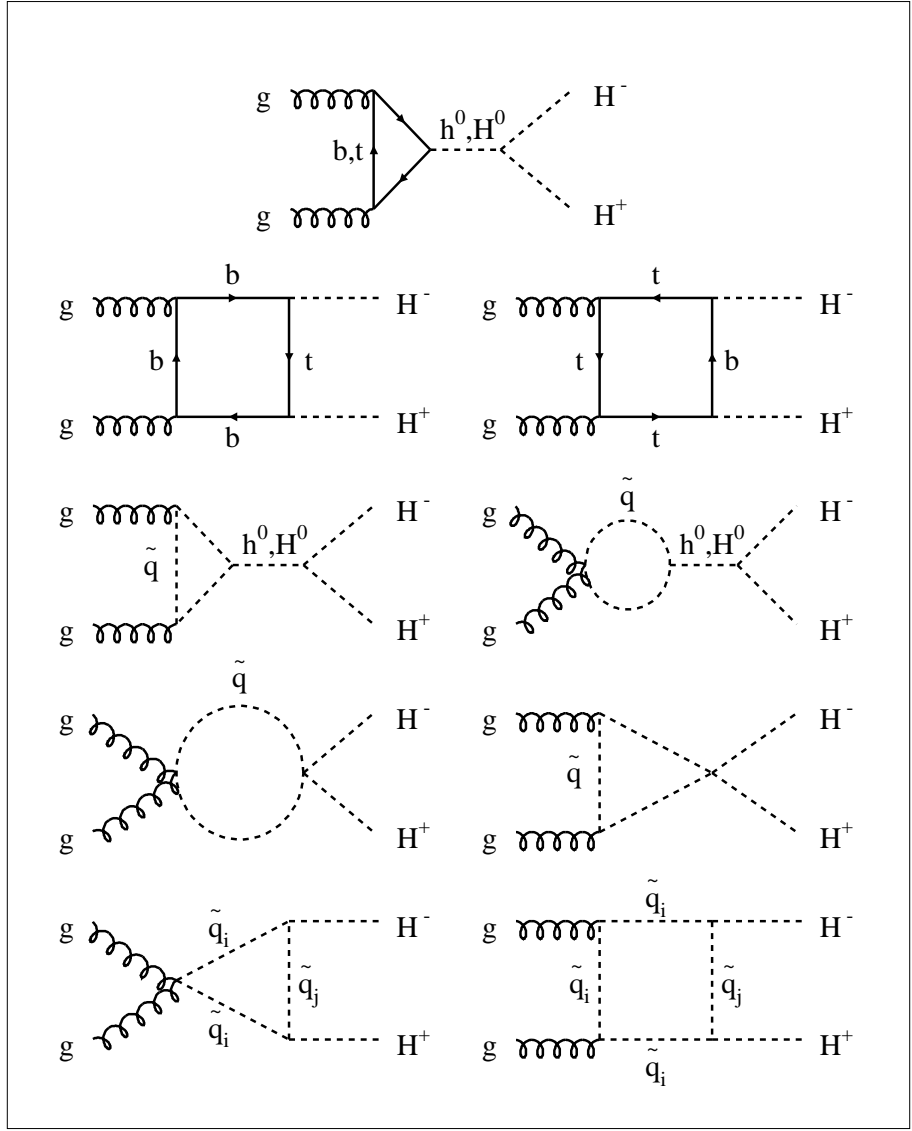,height=18cm}}
\caption{One-loop Feynman diagrams for $gg\to H^+H^-$ in the MSSM.}
\label{fig:two}
\end{center}
\end{figure}

In this paper, we investigate $H^+H^-$ pair production in the MSSM.
At the tree level, this proceeds via $q\bar q$ annihilation,
$q\bar q\to H^+H^-$, where $q=u,d,s,c,b$.
The Drell-Yan process, where a photon and a $Z$-boson are exchanged in the $s$
channel (see upper Feynman diagram in Fig.~\ref{fig:one}) has been studied by
a number of authors \cite{eic}.
As pointed out in Ref.~\cite{wh}, in the case $q=b$, there are additional
Feynman diagrams involving the $h^0$ and $H^0$ bosons in the $s$ channel and
the top quark in the $t$ channel (see middle and lower diagrams in 
Fig.~\ref{fig:one}).
As we shall see later on, for small or large values of $\tan\beta$, their
contribution greatly exceeds the one due to the Drell-Yan process, which is
independent of $\tan\beta$.
To our knowledge, these additional diagrams have not been considered elsewhere 
in the literature.
At one loop, $H^+H^-$ pair production receives an additional contribution from
$gg$ fusion, $gg\to H^+H^-$.
Although the cross section of $gg$ fusion is suppressed by two powers of
$\alpha_s$ relative to the one of $q\bar q$ annihilation, it is expected to
yield a comparable contribution at multi-TeV hadron colliders, due to the
overwhelming gluon luminosity.
In the 2HDM, $gg$ fusion is mediated by heavy-quark loops (see upper two rows
in Fig.~\ref{fig:two}) \cite{wil,kra}.
We calculated these QCD contributions and found full agreement with the 
analytical and numerical results presented in Ref.~\cite{kra}. 
In the MSSM, there are additional QCD contributions induced by squark loops
(see lower three rows in Fig.~\ref{fig:two}), which we shall present here.

As for $b\bar b$ annihilation, it should be noted that the treatment of bottom
as an active flavour inside the colliding hadrons leads to an effective
description, which comprises contributions from the higher-order subprocesses
$gb\to H^+H^- b$, $g\bar b\to H^+H^-\bar b$, and $gg\to H^+H^-b\bar b$.
If all these subprocesses are to be explicitly included along with
$b\bar b\to H^+H^-$, then it is necessary to employ a judiciously subtracted
parton density function (PDF) for the bottom quark in order to avoid double
counting \cite{gun,dad}.
The evaluation of $b\bar b\to H^+H^-$ with an unsubtracted bottom PDF is
expected to slightly overestimate the true cross section \cite{gun,dad}.
For simplicity, we shall nevertheless adopt this effective approach in our
analysis, keeping in mind that a QCD-correction factor below unity is to be 
applied.

The circumstance that the spectrum of states is more than doubled if one 
passes from the SM to the MSSM gives rise to a proliferation of parameters,
which weakens the predictability of the theory.
A canonical method to reduce the number of parameters is to embed the MSSM
into a grand unified theory (GUT), e.g., a suitable supergravity (SUGRA)
model, in such a way that it is recovered in the low-energy limit.
The MSSM thus constrained is described by the following parameters at the GUT 
scale, which come in addition to $\tan\beta$ and $m_A$: the universal sfermion
mass, $m_0$; the universal gaugino mass, $m_{1/2}$; the trilinear
Higgs-sfermion coupling, $A$; the bilinear Higgs coupling, $B$; and the
Higgs-higgsino mass parameter, $\mu$.
The number of parameters can be further reduced by making additional
assumptions.
Unification of the $\tau$-lepton and $b$-quark Yukawa couplings at the GUT 
scale leads to a correlation between $m_t$ and $\tan\beta$.
Furthermore, if the electroweak symmetry is broken radiatively, then $B$ and
$\mu$ are determined up to the sign of $\mu$.
Finally, it turns out that the MSSM parameters are nearly independent of the
value of $A$, as long as $|A|\alt500$~GeV at the GUT scale.
Further details on the SUGRA-inspired MSSM scenario may be found in
Ref.~\cite{kal} and the references cited therein.

This paper is organized as follows.
In Section~2, we shall list analytic results for the tree-level cross section
of $q\bar q\to H^+H^-$, including the Yukawa-enhanced contributions for $q=b$,
and the squark loop contribution to the $gg\to H^+H^-$ amplitude in the MSSM.
The relevant Higgs-squark coupling constants and the squark loop form factors
are relegated to Appendices~A and B, respectively.
In Section~3, we shall present quantitative predictions for the inclusive
cross section of $pp\to H^+H^-+X$ at the LHC adopting the SUGRA-inspired MSSM.
Section~4 contains our conclusions.

\section{Analytic results}

We start by defining the kinematics of the inclusive reaction
$AB\to H^+H^-+X$, where $A$ and $B$ are colliding hadrons, which are taken to
be massless.
Let $\sqrt S$ be the energy of the initial state and $y$ and $p_T$ the
rapidity and transverse momentum of the $H^+$ boson in the centre-of-mass
(c.m.) system of the collision.
By four-momentum conservation, $m_T\cosh y\le\sqrt S/2$, 
where $m_T=\sqrt{m_H^2+p_T^2}$ is the transverse mass of the $H^\pm$ bosons,
with mass $m_H$.
The hadron $A$ is characterized by its PDF's $F_{a/A}(x_a,M_a)$, where $x_a$
is the fraction of the four-momentum of $A$ which is carried by the (massless)
parton $a$ ($p_a=x_a p_A$), $M_a$ is the factorization scale, and similarly
for $B$.
The Mandelstam variables $s=(p_a+p_b)^2$, $t=(p_a-p_{H^+})^2$, and
$u=(p_b-p_{H^+})^2$ at the parton level are thus related to $S$, $y$, and
$p_T$ by $s=x_ax_bS$, $t=m_H^2-x_a\sqrt Sm_T\exp(-y)$, and
$u=m_H^2-x_b\sqrt Sm_T\exp(y)$, respectively.
Notice that $sp_T^2=tu-m_H^4$.
In the parton model, the differential cross section of $AB\to H^+H^-+X$ is
given by
\begin{eqnarray}
\frac{d^2\sigma}{dy\,dp_T^2}(AB\to H^+H^-+X)
&=&\sum_{a,b}\int_{\bar x_a}^1dx_a\,F_{a/A}(x_a,M_a)F_{b/B}(x_b,M_b)
\frac{x_bs}{m_H^2-t}
\nonumber\\
&&{}\times\frac{d\sigma}{dt}(ab\to H^+H^-),
\end{eqnarray}
where $\bar x_a=m_T\exp(y)/[\sqrt S-m_T\exp(-y)]$ and
$x_b=x_am_T\exp(-y)/[x_a\sqrt S-m_T\exp(y)]$.
The parton-level cross section is calculated from the $ab\to H^+H^-$
transition-matrix element ${\cal T}$ as
$d\sigma/dt=\overline{|{\cal T}|^2}/(16\pi s^2)$, where the average is over
the spin and colour degrees of freedom of the partons $a$ and $b$.

We now turn to the specific subprocesses $ab\to H^+H^-$.
We work in the MSSM, adopting the Feynman rules from Ref.~\cite{hab}.
For convenience, we introduce the short-hand notations
$s_w=\sin\theta_w$, $c_w=\cos\theta_w$, $s_\alpha=\sin\alpha$,
$c_\alpha=\cos\alpha$, $s_\beta=\sin\beta$, $c_\beta=\cos\beta$,
$s_{2\beta}=\sin(2\beta)$, $c_{2\beta}=\cos(2\beta)$,
$s_\pm=\sin(\alpha\pm\beta)$, and $c_\pm=\cos(\alpha\pm\beta)$, where
$\theta_w$ is the electroweak mixing angle, $\alpha$ is the mixing angle that
rotates the weak CP-even Higgs eigenstates into the mass eigenstates $h^0$ and
$H^0$, and $\tan\beta=v_2/v_1$ is the ratio of the vacuum expectation values
of the two Higgs doublets.
We neglect the Yukawa couplings of the first- and second-generation quarks.
We treat the $b$ and $\bar b$ quarks as active partons inside the colliding
hadrons $A$ and $B$.
This should be a useful picture at the energies of interest here,
$\sqrt S>2m_H$.
For consistency with the underlying infinite-momentum frame, we neglect the
bottom-quark mass, $m_b$.
However, we must not suppress terms proportional to $m_b$ in the Yukawa
couplings, since they generally dominate the related $m_t$-dependent terms if
$\tan\beta$ is large enough, typically for
$\tan\beta\agt\sqrt{m_t/m_b}\approx6$.
This is obvious for the $H^-\bar bt$ vertex, which has the Feynman rule
\cite{hab}
\begin{equation}
i2^{-1/4}G_F^{1/2}[m_t\cot\beta(1+\gamma_5)+m_b\tan\beta(1-\gamma_5)],
\label{eq:hbt}
\end{equation}
where $G_F$ is Fermi's constant and we have neglected the
Cabibbo-Kobayashi-Maskawa mixing, i.e., $V_{tb}=1$.

The tree-level diagrams for $b\bar b\to H^+H^-$ in the MSSM are depicted in
Fig.~\ref{fig:one}.
The resulting parton-level cross section reads
\begin{equation}
\frac{d\sigma}{dt}(b\bar b\to H^+H^-)=\frac{G_F^2}{3\pi s}
\left[|S|^2+4p_T^2\left(|V+T_+|^2+|A+T_-|^2\right)\right],
\label{eq:bb}
\end{equation}
where
\begin{eqnarray}
S&=&g_{H^+H^-h^0}g_{h^0bb}{\cal P}_{h^0}(s)
+g_{H^+H^-H^0}g_{H^0bb}{\cal P}_{H^0}(s)
-\frac{m_bm_t^2}{2m_W^2\left(m_t^2-t\right)},
\nonumber\\
V&=&g_{H^+H^-Z}v_{Zbb}{\cal P}_{Z}(s)
+g_{H^+H^-\gamma}v_{\gamma bb}{\cal P}_{\gamma}(s),
\nonumber\\
A&=&g_{H^+H^-Z}a_{Zbb}{\cal P}_{Z}(s),
\nonumber\\
T_\pm&=&\frac{m_t^2\cot^2\beta\pm m_b^2\tan^2\beta}
{8m_w^2\left(m_t^2-t\right)},
\end{eqnarray}
with couplings
\begin{eqnarray}
g_{H^+H^-h^0}&=&m_Ws_--\frac{m_Zs_+c_{2\beta}}{2c_w},\qquad
g_{h^0bb}=\frac{m_bs_\alpha}{2m_Wc_\beta},
\nonumber\\
g_{H^+H^-H^0}&=&-m_Wc_-+\frac{m_Zc_+c_{2\beta}}{2c_w},\qquad
g_{H^0bb}=-\frac{m_bc_\alpha}{2m_Wc_\beta},
\nonumber\\
g_{H^+H^-Z}&=&-\frac{c_w^2-s_w^2}{2c_w},\qquad
v_{Zbb}=-\frac{I_b-2s_w^2Q_b}{2c_w},\qquad
a_{Zbb}=-\frac{I_b}{2c_w},
\nonumber\\
g_{H^+H^-\gamma}&=&-s_w,\qquad
v_{\gamma bb}=-s_wQ_b,
\end{eqnarray}
weak isospin $I_b=-1/2$, and electric charge $Q_b=-1/3$.
Here,
\begin{equation}
{\cal P}_{X}(s)=\frac{1}{s-m_X^2+im_X\Gamma_X}
\end{equation}
is the propagator function of particle $X$, with mass $m_X$ and total decay 
width $\Gamma_X$.
We recover the well-known Drell-Yan cross section of $q\bar q\to H^+H^-$
\cite{eic} from Eq.~(\ref{eq:bb}) by putting $S=T_\pm=0$ and substituting
$b\to q$.
The approximation $S=T_\pm=0$ is justified for the quarks of the first and 
second generations, $q=u,d,s,c$, because $S$ and $T_\pm$ are then suppressed
by the smallness of the corresponding Yukawa couplings.
However, the $S$ and $T_\pm$ terms give rise to sizeable contributions in the
case of $q=b$, especially at small or large values of $\tan\beta$.
The full cross section of $q\bar q$ annihilation is obtained by complementing
the $b\bar b$-initiated cross section of Eq.~(\ref{eq:bb}) with the Drell-Yan
cross sections for $q=u,d,s,c$.

The one-loop diagrams for $gg\to H^+H^-$ in the MSSM are displayed in
Fig.~\ref{fig:two}.
As for the quark loops, our analytical results fully agree with those listed
in Ref.~\cite{kra}, and there is no need to repeat them here.
In the squark case, the $\cal T$-matrix elements corresponding to the triangle 
and box diagrams are found to be
\begin{eqnarray}
\tilde{\cal T}_\triangle&=&\frac{G_Fm_W^2}{\sqrt2}\,
\frac{\alpha_s(\mu_r)}{\pi}
\varepsilon_\mu^c(p_a)\varepsilon_\nu^c(p_b)A_1^{\mu\nu}\tilde F_\triangle,
\nonumber\\
\tilde{\cal T}_\Box&=&\frac{G_Fm_W^2}{\sqrt2}\,\frac{\alpha_s(\mu_r)}{\pi}
\varepsilon_\mu^c(p_a)\varepsilon_\nu^c(p_b)
\left(A_1^{\mu\nu}\tilde F_\Box+A_2^{\mu\nu}\tilde G_\Box\right),
\end{eqnarray}
respectively, where $\alpha_s(\mu_r)$ is the strong coupling constant at
renormalization scale $\mu_r$, $\varepsilon_\mu^c(p_a)$ is the polarization
four-vector of gluon $a$ and similarly for gluon $b$, it is summed over the
colour index $c=1,\ldots,8$,
\begin{eqnarray}
A_1^{\mu\nu}&=&g^{\mu\nu}-\frac{2}{s}p_a^\nu p_b^\mu,
\nonumber\\
A_2^{\mu\nu}&=&g^{\mu\nu}+\frac{2}{p_T^2}\left(\frac{m_H^2}{s}p_a^\nu p_b^\mu
+\frac{u-m_H^2}{s}p_a^\nu p_{H^+}^\mu+\frac{t-m_H^2}{s}p_b^\mu p_{H^+}^\nu
+p_{H^+}^\mu p_{H^+}^\nu\right),
\end{eqnarray}
and the form factors $\tilde F_\triangle$, $\tilde F_\Box$, and
$\tilde G_\Box$ are listed in Appendix~B.
Due to Bose symmetry, $\tilde{\cal T}_\triangle$ and $\tilde{\cal T}_\Box$ are
invariant under the simultaneous replacements $\mu\leftrightarrow\nu$ and
$p_a\leftrightarrow p_b$.
Consequently, $\tilde F_\triangle$, $\tilde F_\Box$, and $\tilde G_\Box$ are
symmetric in $t$ and $u$.

The parton-level cross section of $b\bar b\to H^+H^-$ including both quark and 
squark contributions is then given by
\begin{eqnarray}
\frac{d\sigma}{dt}(gg\to H^+H^-)&=&\frac{G_F^2\alpha_s^2(\mu_r)}{256(2\pi)^3}
\left[\left|\sum_{Q=t,b}C_\triangle^QF_\triangle^Q+F_\Box
-\frac{2m_W^2}{s}\left(\tilde F_\triangle+\tilde F_\Box\right)\right|^2
\right.\nonumber\\
&&{}+\left.\left|G_\Box-\frac{2m_W^2}{s}\tilde G_\Box\right|^2
+\left|H_\Box\right|^2\right],
\label{eq:gg}
\end{eqnarray}
where the generalized coupling $C_\triangle^Q$ and the form factors
$F_\triangle^Q$, $F_\Box$, $G_\Box$, and $H_\Box$ may be found in Eq.~(8) and
Appendix~A of Ref.~\cite{kra}, respectively.

\section{Phenomenological implications}

We are now in a position to explore the phenomenological implications of our
results.
The SM input parameters for our numerical analysis are taken to be
$G_F=1.16639\cdot10^{-5}$~GeV$^{-2}$ \cite{pdg}, $m_W=80.394$~GeV,
$m_Z=91.1867$~GeV, $m_t=174.3$~GeV \cite{eww}, and $m_b=4.7$~GeV.
We adopt the lowest-order set CTEQ5L \cite{lai} of proton PDF's.
We evaluate $\alpha_s(\mu_r)$ from the lowest-order formula \cite{pdg} with
$n_f=5$ quark flavours and asymptotic scale parameter
$\Lambda_{\rm QCD}^{(5)}=146$~MeV \cite{lai}.
We identify the renormalization and factorization scales with the $H^+H^-$
invariant mass, $\mu_r^2=M_a^2=M_b^2=s$.
For our purposes, it is useful to replace $m_A$ by $m_H$, the mass of the
$H^\pm$ bosons to be produced, in the set of MSSM input parameters.
We vary $\tan\beta$ and $m_H$ in the ranges $1<\tan\beta<40\approx m_t/m_b$
and 120~GeV${}<m_H<550$~GeV, respectively.
As for the GUT parameters, we choose $m_{1/2}=100$~GeV, $A=0$, and $\mu<0$, 
and tune $m_0$ so as to be consistent with the desired value of $m_H$.
All other MSSM parameters are then determined according to the SUGRA-inspired
scenario as implemented in the program package SUSPECT \cite{djo}.
We do not impose the unification of the $\tau$-lepton and $b$-quark Yukawa
couplings at the GUT scale, which would just constrain the allowed $\tan\beta$
range without any visible effect on the results for these values of
$\tan\beta$.

\begin{figure}[ht]
\begin{center}
\centerline{\epsfig{figure=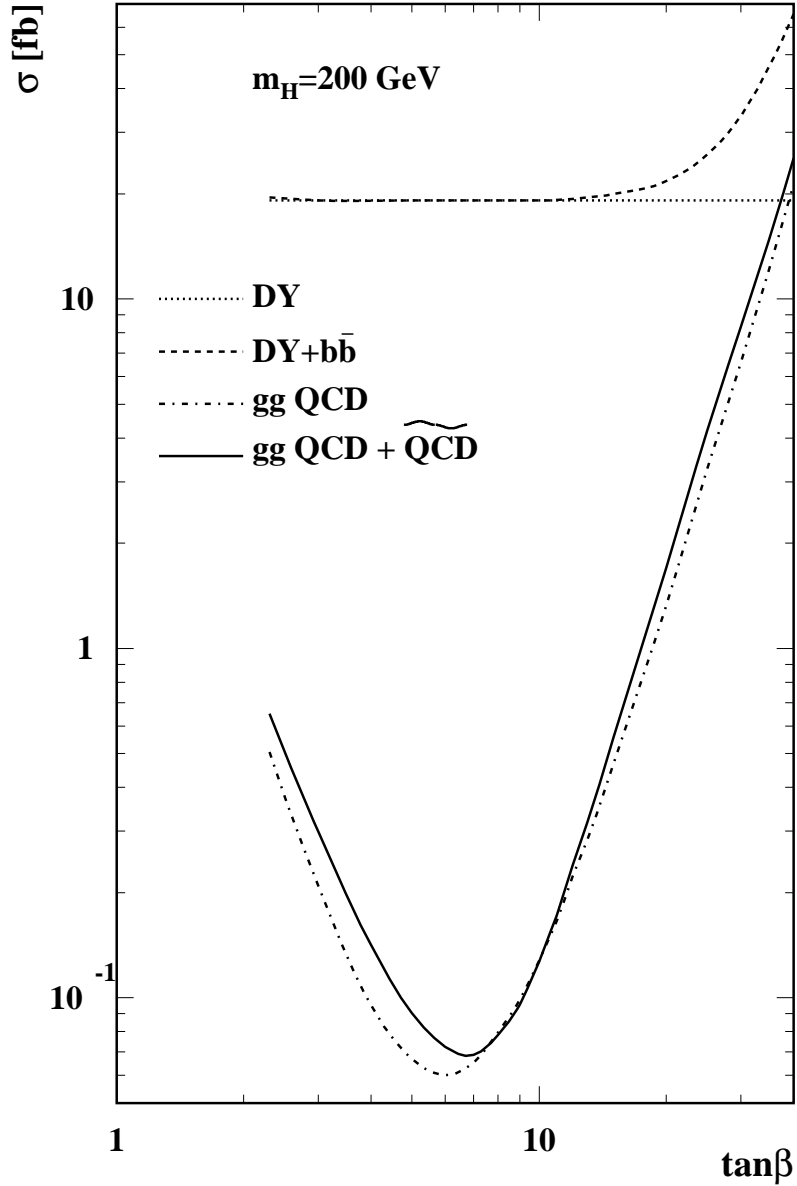,height=18cm}}
\caption{Total cross sections $\sigma$ (in fb) of $pp\to H^+H^-+X$ via
$q\bar q$ annihilation (dashed line) and $gg$ fusion (solid line) at the LHC
as functions of $\tan\beta$ for $m_H=200$~GeV.
For comparison, also the Drell-Yan contribution to $q\bar q$ annihilation 
(dotted line) and the quark loop contribution to $gg$ fusion (dot-dashed line)
are shown.}
\label{fig:three}
\end{center}
\end{figure}

We now study the total cross section of $pp\to H^+H^-+X$ at the LHC, with
c.m.\ energy $\sqrt S=14$~TeV.
In Fig.~\ref{fig:three}, the full contributions due to $q\bar q$ annihilation
(dashed line) and $gg$ fusion (solid line) are displayed as functions of
$\tan\beta$ for $m_H=200$~GeV.
For comparison, also the Drell-Yan contributions to $q\bar q$ annihilation for
$q=u,d,s,c,b$ (dotted line) and the quark loop contribution to $gg$ fusion
(dot-dashed line), which is the full one-loop result for $gg$-fusion in the 
2HDM, are shown.
In the case of $q\bar q$ annihilation, as anticipated in Section~1, the
Yukawa-enhanced contribution for $q=b$ greatly enhances the conventional
Drell-Yan cross section for large values of $\tan\beta$, by more than a factor
of three for $\tan\beta=40$.
The expected enhancement for small $\tan\beta$ is invisible, since solutions
with $\tan\beta\alt2$ are excluded in the SUGRA-inspired MSSM for our choice
of input parameters.
As for $gg$ fusion, the dot-dashed line nicely agrees with Fig.~6 of 
Ref.~\cite{kra}.
The quark loop contribution exhibits a minimum at
$\tan\beta\approx\sqrt{m_t/m_b}\approx6$.
This may be understood by observing that the average strength of the
$H^-\bar bt$ coupling in Eq.~(\ref{eq:hbt}), proportional to
$\sqrt{m_t^2\cot^2\beta+m_b^2\tan^2\beta}$, is then minimal \cite{kra}.
Passing from the 2HDM to the MSSM, we need to coherently add the squark loop
contribution according to Eq.~(\ref{eq:gg}).
We observe that this leads to a significant rise in cross section, by up to
50\%, unless $\tan\beta$ is close to 10.
Nevertheless, the full tree-level cross section is dominant for all values of
$\tan\beta$.

\begin{figure}[ht]
\begin{center}
\centerline{\epsfig{figure=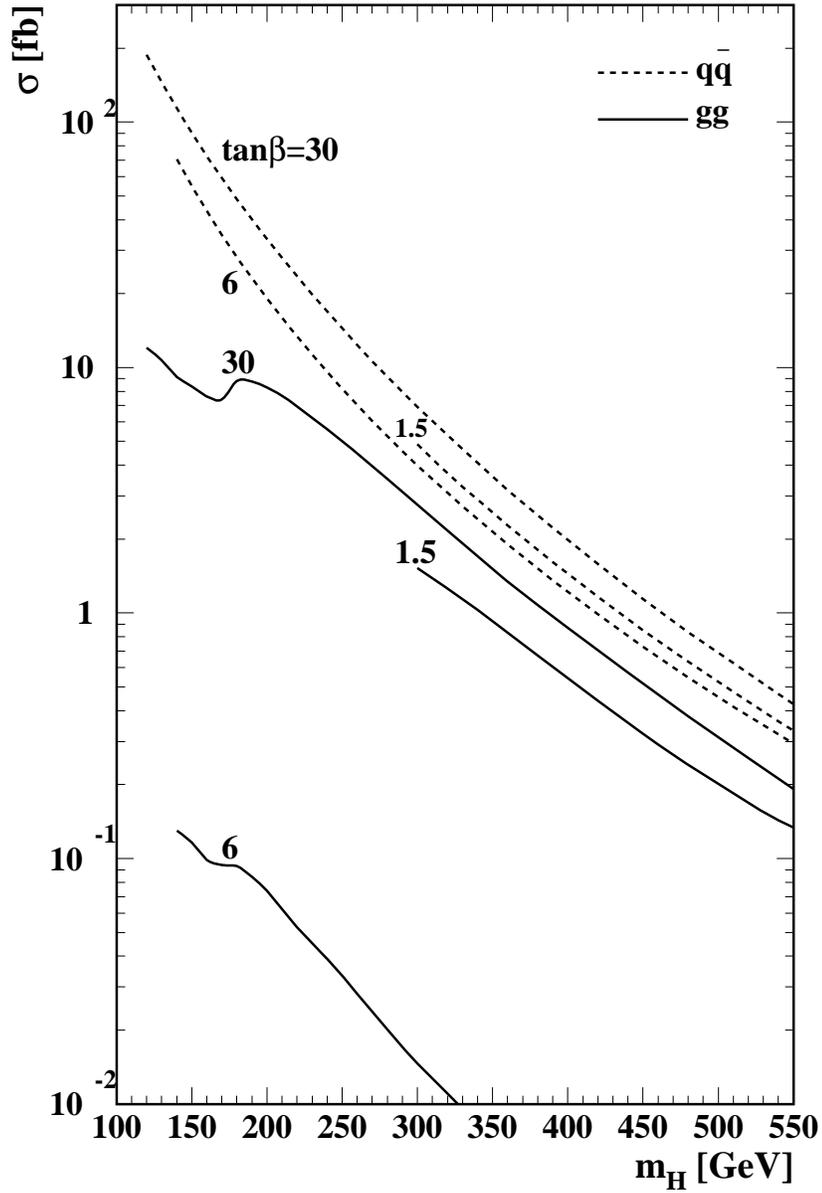,height=18cm}}
\caption{Total cross sections $\sigma$ (in fb) of $pp\to H^+H^-+X$ via
$q\bar q$ annihilation (dashed lines) and $gg$ fusion (solid lines) at the LHC
as functions of $m_H$ for $\tan\beta=1.5$, 6, and 30.}
\label{fig:four}
\end{center}
\end{figure}

In Fig.~\ref{fig:four}, the $m_H$ dependence of the full
$q\bar q$-annihilation (dashed lines) and $gg$-fusion cross sections (solid
lines) is studied for $\tan\beta=1.5$, 6, and 30.
As we have already seen in Fig.~\ref{fig:three}, $q\bar q$ annihilation always
dominates.
Its contribution modestly exceeds the one due to $gg$ fusion, by a factor of
three or less, if $\tan\beta\approx1.5$ or 30 and $m_H\agt200$~GeV, but it is
more than one order of magnitude larger if $m_H\alt m_t$.
The $gg$-fusion contribution is greatly suppressed if $\tan\beta\approx6$,
independently of $m_H$.
For all values of $\tan\beta$, the latter exhibits a dip located about
$m_H=m_t$, which arises from resonating top-quark propagators in the quark box
form factors.
(In the case of $\tan\beta=1.5$, this dip lies in the excluded $m_H$ range.)
This feature may also be seen in Fig.~5 of Ref.~\cite{kra}, where the quark
loop contribution is shown separately.
We note in passing that we also find good agreement with that figure.

For a comparison with future experimental data, the $q\bar q$-annihilation and
$gg$-fusion channels should be combined.
From Fig.~\ref{fig:four}, we read off that the total cross section of
$pp\to H^+H^-+X$ at the LHC is predicted to be 200~fb (1~fb) in the considered
MSSM scenario if $\tan\beta=30$ and $m_H=120$~GeV (500~GeV).
If we assume the integrated luminosity per year to be at its design value of
$L=100$~fb$^{-1}$ for each of the two LHC experiments, ATLAS and CMS, then
this translates into about 40.000 (200) signal events per year.

\section{Conclusions}

We studied the hadroproduction of $H^+H^-$ pairs within the MSSM, adopting a
SUGRA-inspired scenario.
We included the contributions from $q\bar q$ annihilation and $gg$ fusion to
lowest order and provided full analytic results.
Our analysis reaches beyond previous studies \cite{eic,wil,kra} in two
important respects.
In the case of $q\bar q$ annihilation, we demonstrated that previously 
neglected Yukawa-type contributions in the $b\bar b$ channel lead to a
substantial increase in cross section if $\tan\beta$ is large, by more than a 
factor of three for $\tan\beta=40$.
In the case of $gg$ fusion, we upgraded a previous result \cite{kra}, which we
confirmed, from the 2HDM to the MSSM by including the contributions induced by
virtual squarks.
As a result, the $gg$-fusion cross section may be significantly enhanced, by
up to 50\%, depending on $\tan\beta$.
Should the MSSM be realized in nature, then $H^+H^-$ pair production will 
provide a copious source of charged Higgs bosons at the LHC, with an annual
yield of up to 40.000 signal events, which amounts to 80.000 charged Higgs
bosons per year.

\vspace{1cm}
\noindent
{\bf Acknowledgements}
\smallskip

\noindent
B.A.K. thanks the Theory Group of the Werner-Heisenberg-Institut for the
hospitality extended to him during a visit when this paper was finalized.
The work of A.A.B.B. was supported by the Friedrich-Ebert-Stiftung through
Grant No.~219747.

\def\theequation{\Alph{section}.\arabic{equation}}
\begin{appendix}
\setcounter{equation}{0}
\section{Higgs-squark couplings}

In this appendix, we collect the couplings of the $h^0$, $H^0$, and $H^\pm$ 
bosons to the squarks $\tilde q_i$, with $q=t,b$ and $i=1,2$, which are 
relevant for our analysis.
Defining the mixing matrix which rotates the left- and right-handed squark
fields, $\tilde q_L$ and $\tilde q_R$, into the mass eigenstates $\tilde q_i$
as
\begin{equation}
{\cal M}^{\tilde q}=\left(
\begin{array}{cc}
\cos\theta_{\tilde q} & \sin\theta_{\tilde q} \\
-\sin\theta_{\tilde q} & \cos\theta_{\tilde q} \\
\end{array}
\right),
\end{equation}
we have \cite{hab}
\begin{eqnarray}
\left(
\begin{array}{cc}
g_{h^0\tilde t_1\tilde t_1} & g_{h^0\tilde t_1\tilde t_2} \\
g_{h^0\tilde t_2\tilde t_1} & g_{h^0\tilde t_2\tilde t_2} \\
\end{array}
\right) 
&=&{\cal M}^{\tilde t}\left(
\begin{array}{cc}
\frac{m_Zs_+(I_t^3-s_w^2Q_t)}{c_w}-\frac{m_t^2c_\alpha}{m_Ws_\beta} &
-\frac{m_t(\mu s_\alpha+A_tc_\alpha)}{2m_Ws_\beta} \\
-\frac{m_t(\mu s_\alpha+A_tc_\alpha)}{2m_Ws_\beta} &
\frac{m_Zs_+s_w^2Q_t}{c_w}-\frac{m_t^2c_\alpha}{m_Ws_\beta} \\
\end{array}
\right)\left({\cal M}^{\tilde t}\right)^T,
\nonumber\\
\left(
\begin{array}{cc}
g_{h^0\tilde b_1\tilde b_1} & g_{h^0\tilde b_1\tilde b_2} \\
g_{h^0\tilde b_2\tilde b_1} & g_{h^0\tilde b_2\tilde b_2} \\
\end{array}
\right)
&=&{\cal M}^{\tilde b}\left(
\begin{array}{cc}
\frac{m_Zs_+(I^3_b-s_w^2Q_b)}{c_w}+\frac{m_b^2s_\alpha}{m_Wc_\beta} &
\frac{m_b(\mu c_\alpha+A_bs_\alpha)}{2m_Wc_\beta} \\
\frac{m_b(\mu c_\alpha+A_bs_\alpha)}{2m_Wc_\beta} &
\frac{m_Zs_+s_w^2Q_b}{c_w}+\frac{m_b^2s_\alpha}{m_Wc_\beta} \\
\end{array}
\right)\left({\cal M}^{\tilde b}\right)^T,
\nonumber\\
\left(
\begin{array}{cc}
g_{H^0\tilde t_1\tilde t_1} & g_{H^0\tilde t_1\tilde t_2} \\
g_{H^0\tilde t_2\tilde t_1} & g_{H^0\tilde t_2\tilde t_2} \\
\end{array}
\right)
&=&{\cal M}^{\tilde t}\left(
\begin{array}{cc}
-\frac{m_Zc_+(I_t^3-s_w^2Q_t)}{c_W}-\frac{m^2_ts_\alpha}{m_Ws_\beta} &
\frac{m_t(\mu c_\alpha-A_ts_\alpha)}{2m_Ws_\beta} \\
\frac{m_t(\mu c_\alpha-A_ts_\alpha)}{2m_Ws_\beta} &
-\frac{m_Zc_+s_w^2Q_t}{c_w}-\frac{m_t^2s_\alpha}{m_Ws_\beta} \\
\end{array} 
\right)\left({\cal M}^{\tilde t}\right)^T,
\nonumber\\
\left(
\begin{array}{cc}
g_{H^0\tilde b_1\tilde b_1} & g_{H^0\tilde b_1\tilde b_2} \\
g_{H^0\tilde b_2\tilde b_1} & g_{H^0\tilde b_2\tilde b_2} \\
\end{array}
\right) 
&=&{\cal M}^{\tilde b}\left(
\begin{array}{cc}
-\frac{m_Zc_+(I_b^3-s_w^2Q_b)}{c_w}-\frac{m_b^2c_\alpha}{m_Wc_\beta} &
\frac{m_b(\mu s_\alpha-A_bc_\alpha)}{2m_Wc_\beta} \\
\frac{m_b(\mu s_\alpha-A_bc_\alpha)}{2m_Wc_\beta} &
-\frac{m_Zc_+s_w^2Q_b}{c_w}-\frac{m_b^2c_\alpha}{m_Wc_\beta} \\
\end{array}
\right)\left({\cal M}^{\tilde b}\right)^T,
\nonumber\\
\left(
\begin{array}{cc}
g_{H^+\tilde t_1\tilde b_1} & g_{H^+\tilde t_1\tilde b_2} \\
g_{H^+\tilde t_2\tilde b_1} & g_{H^+\tilde t_2\tilde b_2} \\
\end{array}
\right) 
&=&{\cal M}^{\tilde t}\left(
\begin{array}{cc}
\frac{-m_W^2s_{2\beta}+m_t^2\cot\beta+m_b^2\tan\beta}{\sqrt2m_W} &
\frac{m_b(\mu+A_b\tan\beta)}{\sqrt2m_W} \\
\frac{m_t(\mu+A_t\cot\beta)}{\sqrt2m_W} &
\frac{m_tm_b(\tan\beta+\cot\beta)}{\sqrt2m_W} \\
\end{array}
\right)\left({\cal M}^{\tilde b}\right)^T,
\nonumber\\
\left(
\begin{array}{cc}
g_{H^+H^-\tilde t_1\tilde t_1} & g_{H^+H^-\tilde t_1\tilde t_2} \\
g_{H^+H^-\tilde t_2\tilde t_1} & g_{H^+H^-\tilde t_2\tilde t_2} \\
\end{array}
\right) 
&=&{\cal M}^{\tilde t}\left(
\begin{array}{cc}
\frac{c_{2\beta}[I_t^3(1-2c_w^2)-s_w^2Q_t]}{2c_w^2}
-\frac{m_b^2\tan^2\beta}{2m_W^2} &
0 \\
0 &
\frac{c_{2\beta}s_w^2Q_t}{2c_w^2}-\frac{m_t^2\cot^2\beta}{2m_W^2} \\
\end{array}
\right)\left({\cal M}^{\tilde t}\right)^T,
\nonumber\\
\left(
\begin{array}{cc}
g_{H^+H^-\tilde b_1\tilde b_1} & g_{H^+H^-\tilde b_1\tilde b_2} \\
g_{H^+H^-\tilde b_2\tilde b_1} & g_{H^+H^-\tilde b_2\tilde b_2} \\
\end{array}
\right)
&=&{\cal M}^{\tilde b}\left(
\begin{array}{cc}
\frac{c_{2\beta}[I_b^3(1-2c_w^2)-s_w^2Q_b]}{2c_w^2}
-\frac{m_t^2\cot^2\beta}{2m_W^2} &
0 \\
0 &
\frac{c_{2\beta}s_w^2Q_b}{2c_w^2}-\frac{m_b^2\tan^2\beta}{2m_W^2} \\
\end{array}
\right)\left({\cal M}^{\tilde b}\right)^T.
\nonumber\\
\end{eqnarray}
The corresponding Feynman rules emerge by multiplying these couplings with
$ig$, where $g$ is the SU(2) coupling constant.
Similar relations apply for the squarks of the first and second generations,
which are also included in our analysis.
However, in these cases, we neglect terms which are suppressed by the
smallness of the corresponding light-quark masses.

\setcounter{equation}{0}
\section{Squark loop form factors}

In this appendix, we express the squark triangle and box form factors,
$\tilde F_\triangle$, $\tilde F_\Box$, and $\tilde G_\Box$, in terms of the 
standard scalar three- and four-point functions,
\begin{eqnarray}
\lefteqn{C_0\left(p_1^2,(p_2-p_1)^2,p_2^2,m_0^2,m_1^2,m_2^2\right)}
\nonumber\\
&=&\int\frac{d^4q}{i\pi^2}\,\frac{1}{\left(q^2-m_0^2+i\epsilon\right)
\left[(q+p_1)^2-m_1^2+i\epsilon\right]\left[(q+p_2)^2-m_2^2+i\epsilon\right]},
\nonumber\\
\lefteqn{D_0\left(p_1^2,(p_2-p_1)^2,(p_3-p_2)^2,p_3^2,p_2^2,(p_3-p_1)^2,
m_0^2,m_1^2,m_2^2,m_3^2\right)}
\nonumber\\
&=&\int\frac{d^4q}{i\pi^2}\,\frac{1}{\left(q^2-m_0^2+i\epsilon\right)
\left[(q+p_1)^2-m_1^2+i\epsilon\right]
\left[(q+p_2)^2-m_2^2+i\epsilon\right]
\left[(q+p_3)^2-m_3^2+i\epsilon\right]},
\nonumber\\
\end{eqnarray}
which we evaluate numerically with the aid of the program package FF
\cite{old}.
We have
\begin{eqnarray}
\tilde F_{\triangle}&=&
g_{H^+H^-h^0}{\cal P}_{h^0}(s)\left[
g_{h^0\tilde t_1\tilde t_1}F_1\left(s,m_{\tilde t_1}\right)
+g_{h^0\tilde t_2\tilde t_2}F_1\left(s,m_{\tilde t_2}\right)\right.
\nonumber\\
&&{}+\left.g_{h^0\tilde b_1\tilde b_1}F_1\left(s,m_{\tilde b_1}\right)
+g_{h^0\tilde b_2\tilde b_2}F_1\left(s,m_{\tilde b_2}\right)\right]
\nonumber\\
&&{}+g_{H^+H^-H^0}{\cal P}_{H^0}(s)\left[
g_{H^0\tilde t_1\tilde t_1}F_1\left(s,m_{\tilde t_1}\right)
+g_{H^0\tilde t_2\tilde t_2}F_1\left(s,m_{\tilde t_2}\right)\right.
\nonumber\\
&&{}+\left.g_{H^0\tilde b_1\tilde b_1}F_1\left(s,m_{\tilde b_1}\right)
+g_{H^0\tilde b_2\tilde b_2}F_1\left(s,m_{\tilde b_2}\right)\right]
\nonumber\\
&&{}-g_{H^+H^-\tilde t_1\tilde t_1}F_1\left(s,m_{\tilde t_1}\right)
-g_{H^+H^-\tilde t_2\tilde t_2}F_1\left(s,m_{\tilde t_2}\right)
\nonumber\\
&&{}-g_{H^+H^-\tilde b_1\tilde b_1}F_1\left(s,m_{\tilde b_1}\right)
-g_{H^+H^-\tilde b_2\tilde b_2}F_1\left(s,m_{\tilde b_2}\right),
\nonumber\\
\tilde F_{\Box}&=&
g_{H^+\tilde t_1\tilde b_1}^2\left[
F_2\left(h,s,t,u,m_{\tilde t_1},m_{\tilde b_1}\right)
+\left(m_{\tilde t_1}\leftrightarrow m_{\tilde b_1}\right)\right]
\nonumber\\
&&{}+g_{H^+\tilde t_1\tilde b_2}^2\left[
F_2\left(h,s,t,u,m_{\tilde t_1},m_{\tilde b_2}\right)
+\left(m_{\tilde t_1}\leftrightarrow m_{\tilde b_2}\right)\right]
\nonumber\\
&&{}+g_{H^+\tilde t_2\tilde b_1}^2\left[
F_2\left(h,s,t,u,m_{\tilde t_2},m_{\tilde b_1}\right)
+\left(m_{\tilde t_2}\leftrightarrow m_{\tilde b_1}\right)\right]
\nonumber\\
&&{}+g_{H^+\tilde t_2\tilde b_2}^2\left[
F_2\left(h,s,t,u,m_{\tilde t_2},m_{\tilde b_2}\right)
+\left(m_{\tilde t_2}\leftrightarrow m_{\tilde b_2}\right)\right],
\end{eqnarray}
where $h=m_H^2$.
$\tilde G_{\Box}$ is obtained from $\tilde F_{\Box}$ by substituting
$F_2\to F_3$.
Here, we have introduced the auxiliary functions
\begin{eqnarray}
F_1(s,m_{\tilde q})&=&2+4m_{\tilde q}^2 
C_0\left(0,0,s,m_{\tilde q}^2,m_{\tilde q}^2,m_{\tilde q}^2\right),
\nonumber\\
F_2(h,s,t,u,m_{\tilde t},m_{\tilde b})&=&
-\frac{2}{s}\left(t_1C_0^3+u_1C_0^5\right)+p_T^2D_0^1
+2m_{\tilde t}^2\left(D_0^1+D_0^3+D_0^5\right),
\nonumber\\
F_3(h,s,t,u,m_{\tilde t},m_{\tilde b})&=&
\frac{1}{sp_T^2}\left\{s\left[-(t+u)C_0^2+t^2D_0^3+u^2D_0^5\right] 
-2tt_1C_0^4-2uu_1C_0^6\right.
\nonumber\\
&&{}+(t^2+u^2-2h^2)C_0^7
\nonumber\\
&&{}+2m_{\tilde t}^2\left[s\left(C_0^1-C_0^2+p_T^2D_0^1-tD_0^2-uD_0^4\right)
-t_1^2D_0^3-u_1^2D_0^5\right]
\nonumber\\
&&{}+\left.sm_{\tilde t}^2\left(m_{\tilde  t}^2-m_{\tilde  b}^2\right)
\left(2D_0^1+D_0^2+D_0^3+D_0^4+D_0^5\right)\right\},
\end{eqnarray}
where $t_1=t-h$, $u_1=u-h$, and
\begin{eqnarray}
C_0^1&=&C_0\left(0,0,s,m_{\tilde b}^2,m_{\tilde b}^2,m_{\tilde b}^2\right),
\nonumber\\
C_0^2&=&C_0\left(0,0,s,m_{\tilde t}^2,m_{\tilde t}^2,m_{\tilde t}^2\right),
\nonumber\\
C_0^3&=&C_0\left(h,0,t,m_{\tilde b}^2,m_{\tilde t}^2,m_{\tilde t}^2\right),
\nonumber\\
C_0^4&=&C_0\left(h,0,t,m_{\tilde t}^2,m_{\tilde b}^2,m_{\tilde b}^2\right),
\nonumber\\
C_0^5&=&C_0\left(h,0,u,m_{\tilde b}^2,m_{\tilde t}^2,m_{\tilde t}^2\right),
\nonumber\\
C_0^6&=&C_0\left(h,0,u,m_{\tilde t}^2,m_{\tilde b}^2,m_{\tilde b}^2\right),
\nonumber\\
C_0^7&=&C_0\left(h,h,s,m_{\tilde t}^2,m_{\tilde b}^2,m_{\tilde t}^2\right),
\nonumber\\
D_0^1&=&D_0\left(h,0,h,0,t,u,
m_{\tilde b}^2,m_{\tilde t}^2,m_{\tilde t}^2,m_{\tilde b}^2\right),
\nonumber\\
D_0^2&=&D_0\left(h,h,0,0,s,t,
m_{\tilde b}^2,m_{\tilde t}^2,m_{\tilde b}^2,m_{\tilde b}^2\right),
\nonumber\\
D_0^3&=&D_0\left(h,h,0,0,s,t,
m_{\tilde t}^2,m_{\tilde b}^2,m_{\tilde t}^2,m_{\tilde t}^2\right),
\nonumber\\
D_0^4&=&D_0\left(h,h,0,0,s,u,
m_{\tilde b}^2,m_{\tilde t}^2,m_{\tilde b}^2,m_{\tilde b}^2\right),
\nonumber\\
D_0^5&=&D_0\left(h,h,0,0,s,u,
m_{\tilde t}^2,m_{\tilde b}^2,m_{\tilde t}^2,m_{\tilde t}^2\right).
\end{eqnarray}

\end{appendix}

\newpage

\newpage

\newpage

\newpage

\end{document}